\documentclass[iop,apjl]{emulateapj}
\usepackage{natbib}
\usepackage{amsfonts}
\usepackage{amssymb}

\shorttitle{Parametric tension between even and odd multipole data of WMAP power spectrum}
\shortauthors{Jaiseung Kim and et al.}

\begin{document}
\title{Parametric tension between even and odd multipole data of WMAP power spectrum: unaccounted contamination or missing parameters?}
\author{Jaiseung Kim and Pavel Naselsky}
\affil{Niels Bohr Institute \& Discovery Center, Blegdamsvej 17, DK-2100 Copenhagen, Denmark}
\email{jkim@nbi.dk}
\submitted{Accepted for Publication in the Astrophysical Journal Letter} 

\begin{abstract}
There exist power contrast in even and odd multipoles of WMAP power spectrum at low and intermediate multipole range.
This anomaly is explicitly associated with the angular power spectrum, which are heavily used for cosmological model fitting. 
Having noted this, we have investigated whether even(odd) multipole data set is individually consistent with the WMAP 
concordance model. Our investigation shows the WMAP concordance model does not make a good fit for even(odd) multipole data set, which indicates 
parametric tension between even and odd multipole data set.
Noting tension is highest in primordial power spectrum parameters, we have additionally considered a running spectral index, but find tension increases to even a higher level.
We believe these parametric tensions may be indications of unaccounted contamination or imperfection of the model.
\end{abstract}


\section{Introduction}
For the past years, there have been great successes in measurement of CMB anisotropy by ground and satellite observations  \citep{ACBAR,QUaD1,WMAP5:powerspectra,QUaD:instrument,WMAP5:basic_result,WMAP5:parameter,QUaD2,ACBAR2008,WMAP7:powerspectra,WMAP7:basic_result}.
By comparing the angular power spectrum of the CMB anisotropy with theoretical predictions, we may impose strong constraints on cosmological models \citep{Inflation,Modern_Cosmology,Foundations_Cosmology,Cosmology}.
In spite of remarkable goodness of fit \citep{WMAP5:Cosmology,WMAP7:Cosmology}, there are some features of WMAP data, which are not well explained by the WMAP concordance model \citep{Chiang_NG,Tegmark:Alignment,Multipole_Vector1,Hemispherical_asymmetry,cold_spot1,Universe_odd,alfven,lowl,fnl_power,odd,odd_origin,WMAP7:anomaly,lowl_anomalies}. 
In particular, the power contrast anomaly between even and odd multipoles is explicitly associated with the angular power spectrum, which are mainly used to fit cosmological models \citep{Universe_odd,odd,odd_origin,odd_bolpol,WMAP7:anomaly}.
Having noted this, we have investigated whether even(odd) multipole data set is consistent with the WMAP concordance model. 
Our investigation shows there exist some level of tension, which may be an indication of unaccounted contamination or missing ingredients in the assumed parametric model such as the flat $\Lambda$CDM model.

\section{even(odd) multipole data and cosmological model fitting}
\label{cosmomc}
We may consider CMB anisotropy as the sum of even and odd parity functions:
\begin{eqnarray} 
T(\hat{\mathbf n})=T^+(\hat{\mathbf n})+T^-(\hat{\mathbf n}),  
\end{eqnarray}
where
\begin{eqnarray} 
T^+(\hat{\mathbf n})&=&\frac{T(\hat{\mathbf n})+T(-\hat{\mathbf n})}{2},\\
T^-(\hat{\mathbf n})&=&\frac{T(\hat{\mathbf n})-T(-\hat{\mathbf n})}{2}.
\end{eqnarray}
Using the parity property of spherical harmonics $Y_{lm}(\hat{\mathbf n})=(-1)^l\,Y_{lm}(-\hat{\mathbf n})$ \citep{Arfken},
it is straightforward to show
\begin{eqnarray} 
T^+(\hat{\mathbf n})&=&\sum_{l=\mathrm{even}}\sum_m a_{lm}\,Y_{lm}(\hat{\mathbf n}), \label{T_even}\\
T^-(\hat{\mathbf n})&=&\sum_{l=\mathrm{odd}}\sum_m a_{lm}\,Y_{lm}(\hat{\mathbf n}). \label{T_odd}
\end{eqnarray}
Obviously, the power spectrum of even and odd multipoles are associated with $T^{+}(\hat{\mathbf n})$ and $T^{-}(\hat{\mathbf n})$ respectively.
Given the $\Lambda$CDM model, we do not expect any features distinct between even and odd multipoles.
However, there have been reported power contrast between even and odd multipoles of WMAP $TT$ power spectrum \citep{Universe_odd,odd,odd_origin,odd_bolpol,WMAP7:anomaly}.
At lowest multipoles ($2\le l\le 22$), there is odd multipole preference (i.e. power excess in odd multipoles and deficit in even multipoles) \citep{Universe_odd,odd,odd_origin,odd_bolpol}, and even multipole preference at intermediate multipoles ($200\le l \le400$) \citep{WMAP7:anomaly}.
Additionally, we have investigated TE correlation, and noticed odd multipole preference at ($100\lesssim l\lesssim 200$) and even multipole preference at ($200\lesssim l \lesssim400$), though its statistical significance is not high enough, due to low Signal-to-Noise Ratio of polarization data.
Not surprisingly, these power contrast anomalies are explicitly associated with the angular power spectrum data, which are mainly used to fit cosmological models. 
Having noted this, we have investigated whether the even(odd) multipole data set is consistent with the concordance model.
\begin{figure}[htb!]
\centering
\includegraphics[scale=.55]{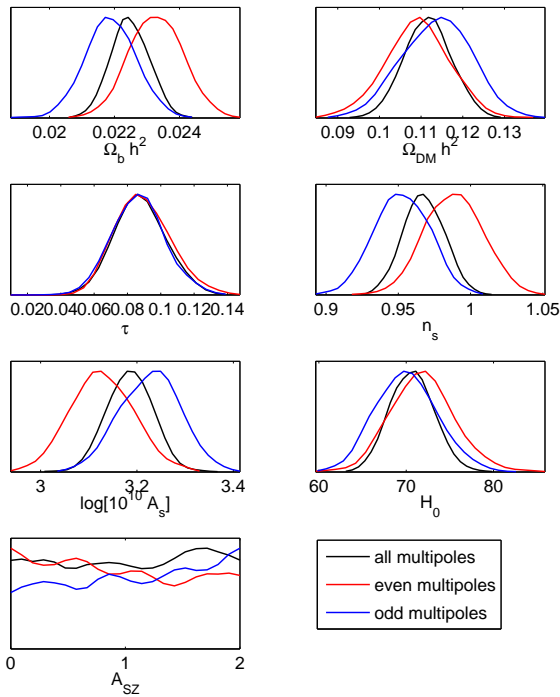}
\caption{Marginalized likelihood of cosmological parameters ($\Lambda$CDM + sz + lens), given whole or even(odd) multipole data.}
\label{like1}
\end{figure}
For a cosmological model, we have considered $\Lambda$CDM + SZ effect + weak-lensing, where cosmological parameters are $\lambda \in \left\{\Omega_b,\Omega_{c},\tau,n_s, A_s, A_{sz}, H_0 \right\}$. For data constraints, we have used the WMAP 7 year TT  and TE power spectrum data, which have been estimated from the ILC map, and cut-sky V and W band maps \citep{WMAP7:powerspectra}. It should be noted that we have used the WMAP power spectrum via the WMAP team's likelihood code. 

In order to use only even(odd) multipole data, we have made slight modifications to the WMAP team's likelihood code, where the Blackwell-Rao estimator 
and the MASTER pseudo-Cl estimator are modified respectively for low multipoles ($l\le32$) and high multipoles ($l>32$) \citep{Gibbs_power,pseudo_Cl,Master_power,WMAP7:powerspectra}.
Note that we have also re-derived the Fisher matrix in accordance with even(odd) multipole data subset.
Additionally, the beam and point source corrections are modified accordingly.
Hereafter, we shall denote WMAP CMB data of whole, even and odd multipoles by $D_0$, $D_2$ and $D_3$ respectively.
Note that even/odd multipole splitting are made for TT and TE power spectrum up to the multipoles of WMAP sensitivity (i.e. $l\le 1200$ for TT and $l\le 800$ for TE).
Using \texttt{CosmoMC} with the modified WMAP likelihood code, we have explored the parameter space on a MPI  cluster with 6 chains \citep{CosmoMC,Gibbs_power,CosmoMC_note,WMAP7:powerspectra}.
For the convergence criterion, we have adopted the Gelman and Rubin's ``variance of chain means'' and set the R-1 statistic to $0.03$ for stopping criterion \citep{Gelman:inference,Gelman:R1}.

In Fig. \ref{like1}, we show the marginalized likelihood of parameters, which are obtained from the run of a \texttt{CosmoMC} with $D_0$, $D_2$ and $D_3$ respectively.
\begin{table}[htb!]
\centering
\caption{cosmological parameters ($\Lambda$CDM + sz + lens)}
\begin{tabular}{cccc}
\hline\hline 
 & $\lambda_0$  &$\lambda_2$   & $\lambda_3$ \\
\hline
$\Omega_{b}\,h^2$  & $0.0226\pm 0.0006$ &$0.0231\pm0.0008$ & $0.0217\pm0.0008$ \\ 
$\Omega_{c}\,h^2$  & $0.112\pm0.006$ &$0.109\pm0.008$ & $0.115\pm0.008$ \\ 
$\tau$  & $0.0837\pm 0.0147$ &$0.0913\pm0.0157$ & $0.0859\pm0.015$ \\ 
$n_s$ & $0.964\pm 0.014$ &$0.989\pm0.02$ & $0.949\pm0.019$ \\ 
$\log[10^{10} A_s]$  & $3.185\pm0.047$ &$3.132\pm0.065$ & $3.239\pm0.062$ \\ 
$H_0$  & $70.53\pm2.48$ &$71.73\pm3.59$ & $69.68\pm3.47$ \\
$A_{\mathrm{sz}}$  & $1.891^{+0.109}_{-1.891}$ &$0.169^{+1.831}_{-0.169}$ & $0.89^{+1.11}_{-0.89}$ \\ 
\hline
\end{tabular}
\label{parameter1}
\end{table}
In Table \ref{parameter1}, we show the best-fit parameters and 1 $\sigma$ confidence intervals, where $\lambda_2$ and $\lambda_3$ denote the best-fit values of $D_2$ and $D_3$ respectively. The parameter set $\lambda_0$ are the best-fit values of whole data $D_0$, and accordingly corresponds to the WMAP concordance model.
As shown in Fig. \ref{like1} and Table \ref{parameter1}, we find non-negligible tension especially in parameters of primordial power spectrum.
It is worth to note that the best-fit spectral index of even multipole data (i.e. $D_2$) is close to a flat spectrum (i.e. $n_s=1$),  while the result from the whole data rule out the flat spectrum by more than 2$\sigma$.

There is a likelihood-ratio test, which allows us to determine the rejection region of an alternative hypothesis, given a null hypothesis \citep{statistics_theory,theoretical_statistics,Statistics_Lupton,Math_methods}.
By setting sets of parameters to a null hypothesis and an alternative hypothesis, we may investigate whether two sets of parameters are consistent with each other.
Therefore, we have evaluated the following in order to assess parametric tension:
\begin{eqnarray*} 
\frac{\mathcal L(\lambda_j|D_i)}{\mathcal L(\lambda_i|D_i)},
\end{eqnarray*}
where parameter set $\lambda_i$ and $\lambda_j$ correspond to a null hypothesis and an alternative hypothesis respectively. 
\begin{table}[htb!]
\centering
\caption{the likelihood ratio: $\Lambda$CDM + sz + lens}
\begin{tabular}{c|ccc}
\hline\hline
 & $\mathcal L(\lambda_0|D_0)$  & $\mathcal L(\lambda_2|D_0)$    & $\mathcal L(\lambda_3|D_0)$  \\
\hline $\mathcal L(\lambda_0|D_0)$  & $1$ &$ 0.076$ & $0.0099$\\
\hline\hline 
& $\mathcal L(\lambda_0|D_2)$  & $\mathcal L(\lambda_2|D_2)$    & $\mathcal L(\lambda_3|D_2)$  \\
\hline
$\mathcal L(\lambda_2|D_2)$  & $0.16$ &$1$ & $2\times 10^{-4}$ \\ 
\hline\hline
& $\mathcal L(\lambda_0|D_3)$  & $\mathcal L(\lambda_2|D_3)$    & $\mathcal L(\lambda_3|D_3)$  \\
\hline
$\mathcal L(\lambda_3|D_3)$   & $0.16$ &$0.0022$ & $1$ \\ 
\hline
\end{tabular}
\label{fit1}
\end{table}
In Table \ref{fit1}, we show the likelihood ratio, where the quantities used for the numerator and denominator are indicated in the uppermost row and leftmost column.
As shown by $\mathcal L(\lambda_0|D_2)/\mathcal L(\lambda_2|D_2)$ and $\mathcal L(\lambda_0|D_3)/\mathcal L(\lambda_3|D_3)$, the WMAP concordance model (i.e. $\lambda_0$) does not make a good fit for even(odd) multipole data set.
Besides, there exist significant tension between two data subsets, as indicated by very small values of 
$\mathcal L(\lambda_3|D_2)/\mathcal L(\lambda_2|D_2)$ and $\mathcal L(\lambda_2|D_3)/\mathcal L(\lambda_3|D_3)$.
The parameter likelihood, except for $A_{\mathrm{sz}}$, follows the shape of Gaussian functions, as shown in Fig. \ref{like1}.
For a likelihood of Gaussian shape, the likelihood ratio 0.1353 and 0.0111 correspond to 2$\sigma$ and 3$\sigma$ significance level respectively.
From Table \ref{fit1}, we may see most of the ratio indicates $\sim 2\sigma$ tension or even higher. 

\begin{figure}[htb!]
\centering
\includegraphics[scale=.55]{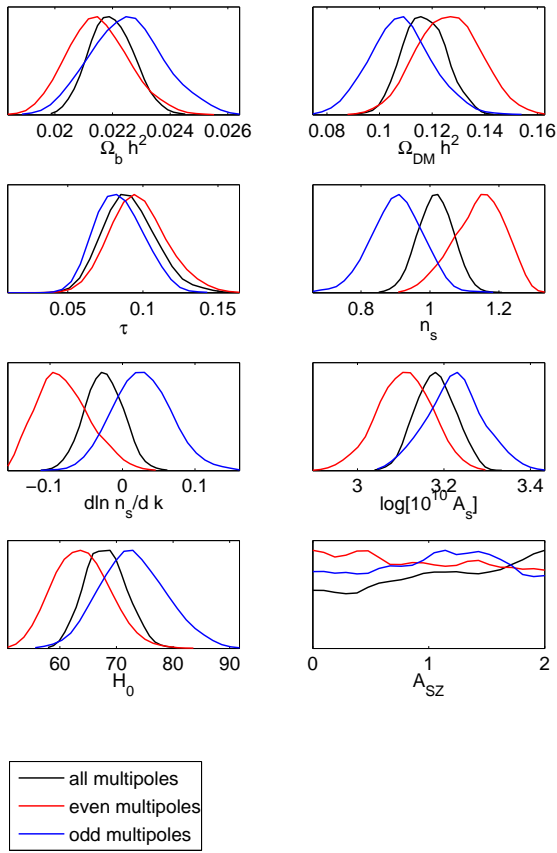}
\caption{Marginalized likelihood of cosmological parameters ($\Lambda$CDM + sz + lens + run), given whole, even and odd multipole data respectively.}
\label{like2}
\end{figure}

\begin{table}[htb!]
\centering
\caption{the likelihood ratio: $\Lambda$CDM + sz + lens + run}
\begin{tabular}{c|ccc}
\hline\hline
 & $\mathcal L(\lambda_0|D_0)$  & $\mathcal L(\lambda_2|D_0)$    & $\mathcal L(\lambda_3|D_0)$  \\
\hline $\mathcal L(\lambda_0|D_0)$  & $1$ & $3.5\times 10^{-4}$ & $0.0078$\\
\hline\hline 
& $\mathcal L(\lambda_0|D_2)$  & $\mathcal L(\lambda_2|D_2)$    & $\mathcal L(\lambda_3|D_2)$  \\
\hline
$\mathcal L(\lambda_2|D_2)$  & $0.06$ &$1$ & $2.3\times 10^{-5}$ \\ 
\hline\hline
& $\mathcal L(\lambda_0|D_3)$  & $\mathcal L(\lambda_2|D_3)$    & $\mathcal L(\lambda_3|D_3)$  \\
\hline
$\mathcal L(\lambda_3|D_3)$   & $0.042$ &$5.8\times 10^{-7}$ & $1$ \\ 
\hline
\end{tabular}
\label{fit2}
\end{table}

As discussed previously, the tension is highest in parameters of primordial power spectrum, which may be an indication of missing parameters in primordial power spectrum (e.g. a running spectral index). Therefore, we have additionally considered a running spectral index $dn_s/d \ln k$ and repeated our investigation.
Surprisingly, we find tension increases to even a higher level. 
We show the marginalized parameter likelihoods and the likelihood ratios in Fig. \ref{like2} and Table \ref{fit2}, where we find tension is also highest in the primordial power spectrum parameters. 

\section{Discussion}
\label{discussion}
The WMAP power contrast anomaly between even and odd multipoles is explicitly associated with the angular power spectrum data, which are mainly used to fit a cosmological model.  
Having noted this, we have investigated whether even(odd) low multipole data set is consistent with the WMAP concordance model. 
Our investigation shows there exists some level of tension. 
Noting tension is highest in primordial power spectrum parameters, we have additionally considered the running of a spectral index $dn_s/d \ln k$, but find tension increases to even a higher level. These parametric tensions may be indications of unaccounted contamination or missing ingredients of the assumed parametric model (i.e. the flat $\Lambda$CDM with/without a running spectral index). Therefore, we believe these parametric tension deserve further investigation.
The Planck surveyor data, which possesses wide frequency coverage and systematics distinct from the WMAP, may allow us to resolve this tension. 
 
\section{Acknowledgments}
We are grateful to the anonymous referee for thorough reading and helpful comments, which lead to significant improvement of this letter.
We are grateful to Hiranya Peiris, Hael Collins, Savvas Nesseris and Wen Zhao for useful discussion.
We acknowledge the use of the Legacy Archive for Microwave Background Data Analysis (LAMBDA). 
Our data analysis made the use of the \texttt{CosmoMC} package \citep{CAMB,CosmoMC}.
This work is supported in part by Danmarks Grundforskningsfond, which allowed the establishment of the Danish Discovery Center.
This work is supported by FNU grant 272-06-0417, 272-07-0528 and 21-04-0355.

\end{document}